\documentclass{iau}
\usepackage{graphicx}

\title[IAUS 289.~~Advancing the physics of cosmic distances] %% give here short title %%
{Visualization of structures and cosmic~flows in the Local Universe}

\author[Daniel Pomar\`ede, H\'el\`ene Courtois \& R. Brent Tully]   %% give here short author list %%
{Daniel Pomar\`ede$^1$, H\'el\`ene Courtois$^{2,3}$ \and  R. Brent Tully$^3$ }

\affiliation{$^1$CEA/Saclay - Irfu, 91191 Gif-sur-Yvette Cedex, France \\ 
$^2$University of Lyon; UCB Lyon 1/CNRS/IN2P3; IPN Lyon, France \\ 
$^3$Institute for Astronomy (IFA), University of Hawaii, 2680 Woodlawn Drive, HI 96822, USA}

\pubyear{2012}
\volume{289}  %% insert here IAU Symposium No.
\pagerange{119--126}
\jname{Advancing the physics of cosmic distances}
\editors{R. de Grijs, G. Bono}
\begin{document}

\maketitle

\begin{abstract}
A visualization of three-dimensional structures and cosmic flows is presented using information from the
Extragalactic Distance Database V8k redshift catalog and peculiar velocities from the Cosmicflows-1 survey.
Structures within a volume bounded at 8000 km/s on the cardinal Supergalactic axes are explored in terms
of both the display of the positions of the 30124 galaxies of the catalog and its reconstructed luminosity density
field, corrected to account for growing incompleteness with distance.  Cosmography of the Local
Universe is discussed with the intent to identify the most
prominent structures, including voids, galaxy clusters, filaments and walls. The mapping also benefits from precise distance measures
provided through the Cosmicflows-1 observational program. 
Three-dimensional visualizations of the
coherent flows of galaxies in the nearby universe are presented using recent results obtained on the reconstruction of cosmic flows with the Wiener Filter approach. The three major components of the Milky
Way motion, namely the expulsion from the Local Void, the infall toward the Virgo Cluster, and the bulk
flow of the historic Local Supercluster toward the Great Attractor are illustrated using different visualization
techniques and analyzed in the light of the cosmography derived from the V8k redshift and Cosmicflows-1 distance catalogs.
\keywords{atlases; distances and redshifts; large scale structures of universe}
%% add here a maximum of 10 keywords, to be taken form the file <Keywords.txt>
\end{abstract}

\firstsection % if your document starts with a section,
              % remove some space above using this command.
\section{Introduction}

The visualization of three-dimensional structures and cosmic flows is a critical ingredient of research into the cosmography of the Local Universe. The purpose of cosmography is to characterize the morphology of
features pertaining to the hierarchy of cosmological structures such as voids, groups, clouds, sheets,
clusters, filaments, chains, superclusters, and walls. Since structures at all scales are not static in
the Hubble flow, cosmography also has to deal with kinematic information. 

Maps that assume relative distances based on galaxy redshifts are distorted from true 3D positions. The analysis of cosmic flows in the context of the spatial distribution of
charted structures involves identification of attractors and other sources of motions. In turn,
the reconstruction of flows provides new insights on their source density field. In this context,
accurate determinations of distances and the peculiar velocity of galaxies are required.

\vspace{-.6 cm}
\section{The COSMIC FLOWS Project}

The objective of the COSMIC FLOWS Project is to improve our knowledge of the cosmography
of the nearby universe, with an emphasis on the identification of attractors.
A key issue is to measure the radial component of the galaxies deviant motion from the Hubble cosmic expansion.
The radial peculiar velocity $V_{pec}$ is expressed as a function of the velocity relative to the CMB background, the
distance $d$ and the value of the Hubble Constant ${\rm H}_{0}$ : $V_{pec}=V_{CMB}-d* {\rm H}_{0}$.
Within 10 Mpc, the Cepheid Period-Luminosity relation,
the Tip of the Red Giant Branch method and the Surface Brightness Fluctuation luminosity indicators provide
measurements of the distances with 10\% accuracy. Further up to 200 Mpc, the galaxy neutral HI gas
luminosity-rotation rate Tully-Fisher correlation gives distance measurements with an accuracy of $\sim$ 20\%.
HI observations most recently have been conducted using the NRAO 100-meter diameter Robert C. Byrd Green Bank Telescope, complemented in the
southern sky with the Parkes Telescope. The photometry is acquired
with several systems including the University of Hawaii 2.2m Telescope.
A first release of measurements of distances and peculiar velocities of 1797 galaxies was achieved with the
{\sl Cosmicflows-1} Catalog presented in \cite[Tully \etal\ (2008)]{2008ApJ...676..184T}.
A three-dimensional velocity and density reconstruction of these peculiar velocities using the Wiener Filter technique
%and highlighting the influence of the ``Great Attractor" in the direction of Centaurus
was presented in \cite[Courtois \etal\ (2012)]{2012ApJ...744...43C}. 

\vspace{-.6 cm}
\section{The Extragalactic Distance Database Catalogs}

Our study of the cosmography of the Local Universe benefits from the catalogs made available through the
Extragalactic Distance Database presented in \cite[Tully \etal\ (2009)]{2009AJ....138..323T}. Besides the 
{\sl Cosmicflows-1} Catalog, the {\sl V8k} Catalog consists in 30124 galaxies with redshift found within
8000 km/s along each axis of the supergalactic coordinate system defined
by \cite[de Vaucouleurs \etal\ (1991)]{devaucouleur1991}. To account
for the distortions caused by the influence of the Virgo cluster, redshift positions within 3000 km/s are adjusted
following a numerical action flow model. Redshift distortions associated with
virial velocities within clusters are also corrected. 
%These corrections are of primary importance
%in the context of cosmography where the objective is to present undistorted maps of structures.

A weakness in the use of these catalogs consists in their growing incompleteness with
distance. A description that
takes this bias into account is offered by the reconstruction
of the luminosity density field obtained by smoothing the distribution of galaxies corrected with a
Schechter function, as described in \cite[Courtois \etal\ (in preparation)]{inprep}.

\vspace{-.6 cm}
\section{Cosmographical maps of the Local Universe}

\begin{figure}[ht]
\begin{center}
 \includegraphics[width=0.73\textwidth]{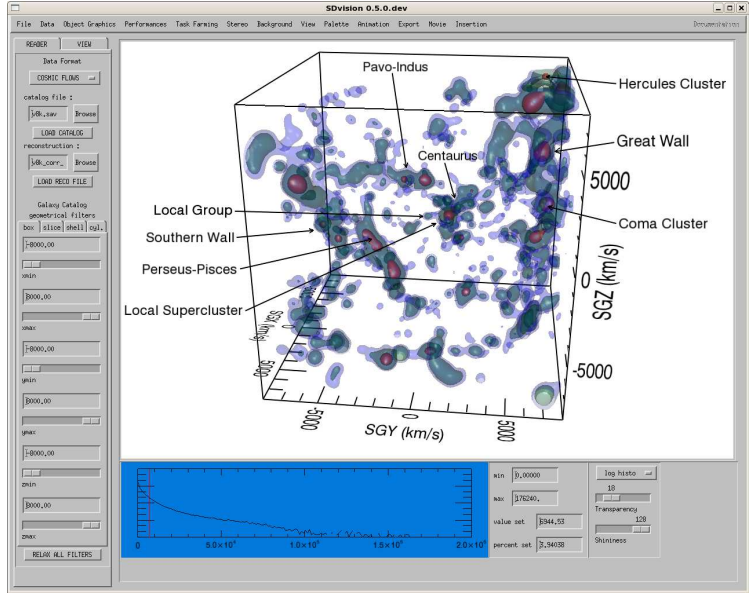} 
 \caption{Visualization of the {\sl V8k} Catalog smoothed density field with SDvision using multiple
 semi-transparent isosurfaces and providing some elements of cosmography.}
   \label{widget}
\end{center}
\end{figure}

Cosmography calls upon the tools of visualization. An assembly of individual galaxies forms a cloud of points, each one
possibly attached with a pointing velocity vector. Reconstructed density fields are described on
uniform grids, visualized with several techniques such as isosurface reconstruction.
Velocity fields are visualized using vectors or streamlines. 
To address these needs, the SDvision interactive visualization software is designed to explore structures and cosmic flows in
three dimensions, and produce maps and exploratory
movies, as illustrated in Fig.\,\ref{widget}.

\begin{figure}[ht]
\begin{center}
\vspace{-.3 cm}
 \includegraphics[width=0.48\textwidth]{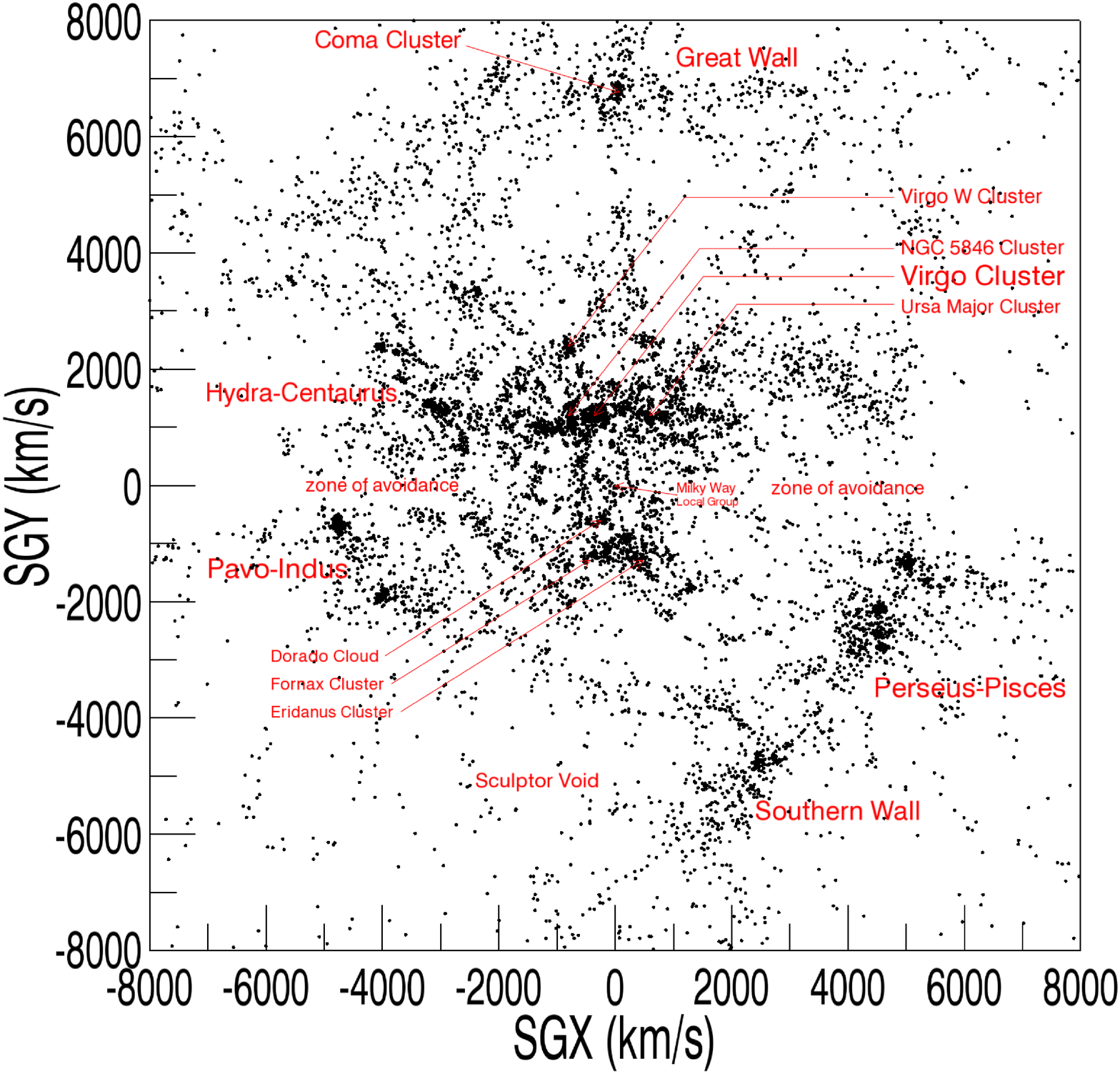} 
 \includegraphics[width=0.48\textwidth]{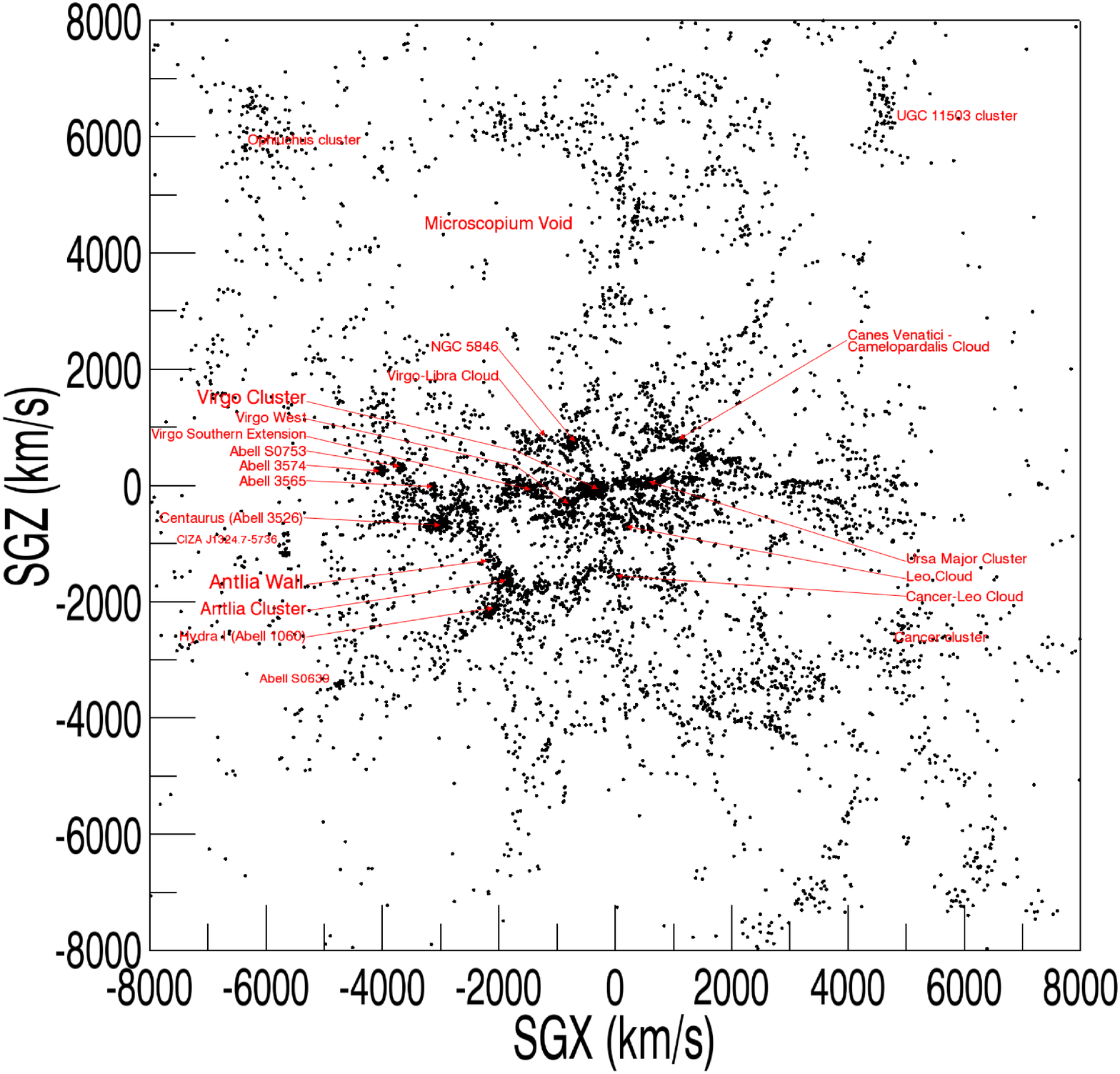} 
\vspace{-.3 cm}
 \caption{Slices of the Local Universe. These two projections display the distributions of galaxies 
 for $-1000\leq$ SGZ (km/s) $\leq 1000$ (left) and $500\leq$ SGY (km/s) $\leq 2500$ (right).}
   \label{slices}
\end{center}
\end{figure}

\begin{figure}[ht]
\begin{center}
 \includegraphics[width=0.46\textwidth]{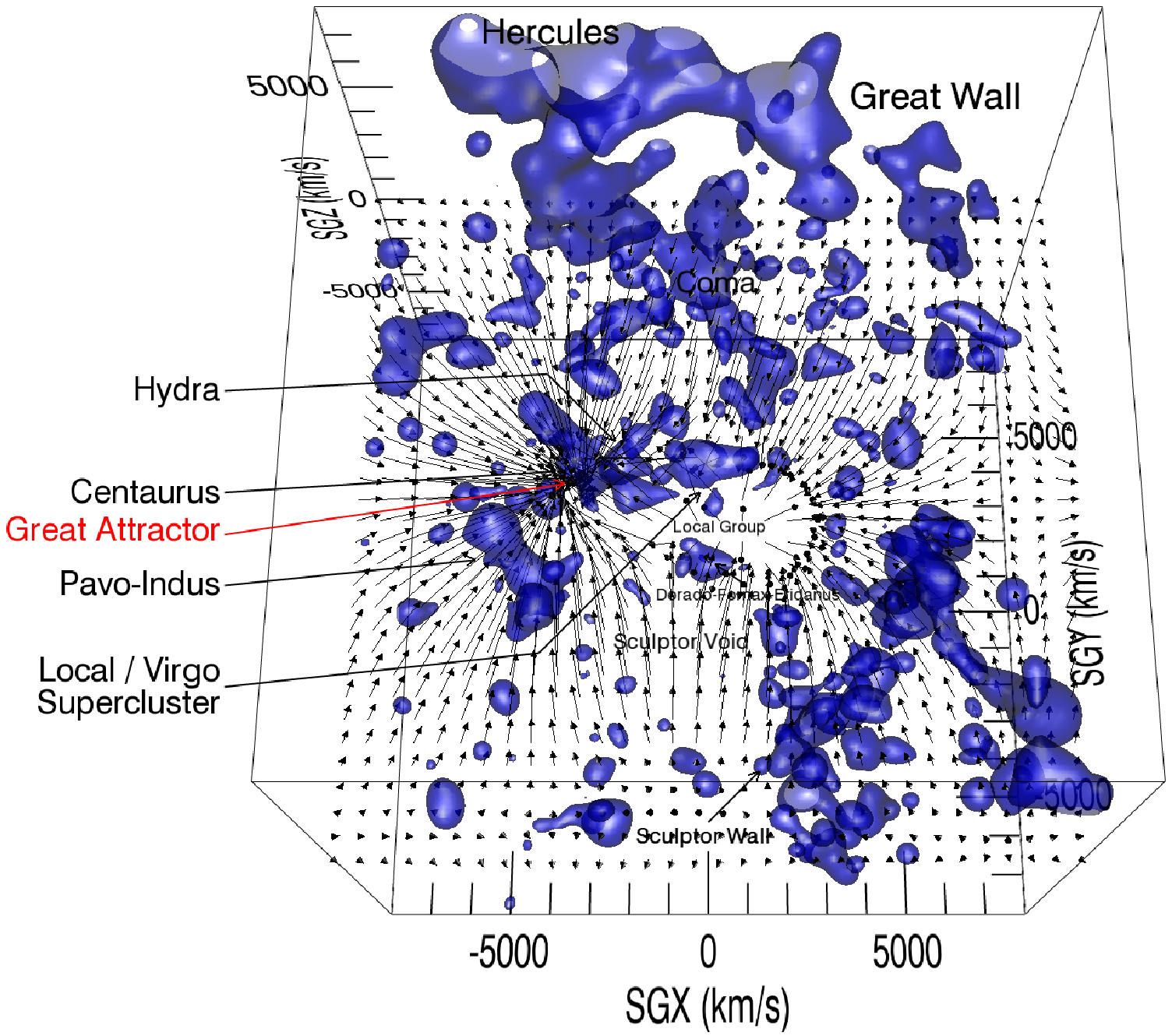} 
 \includegraphics[width=0.39\textwidth]{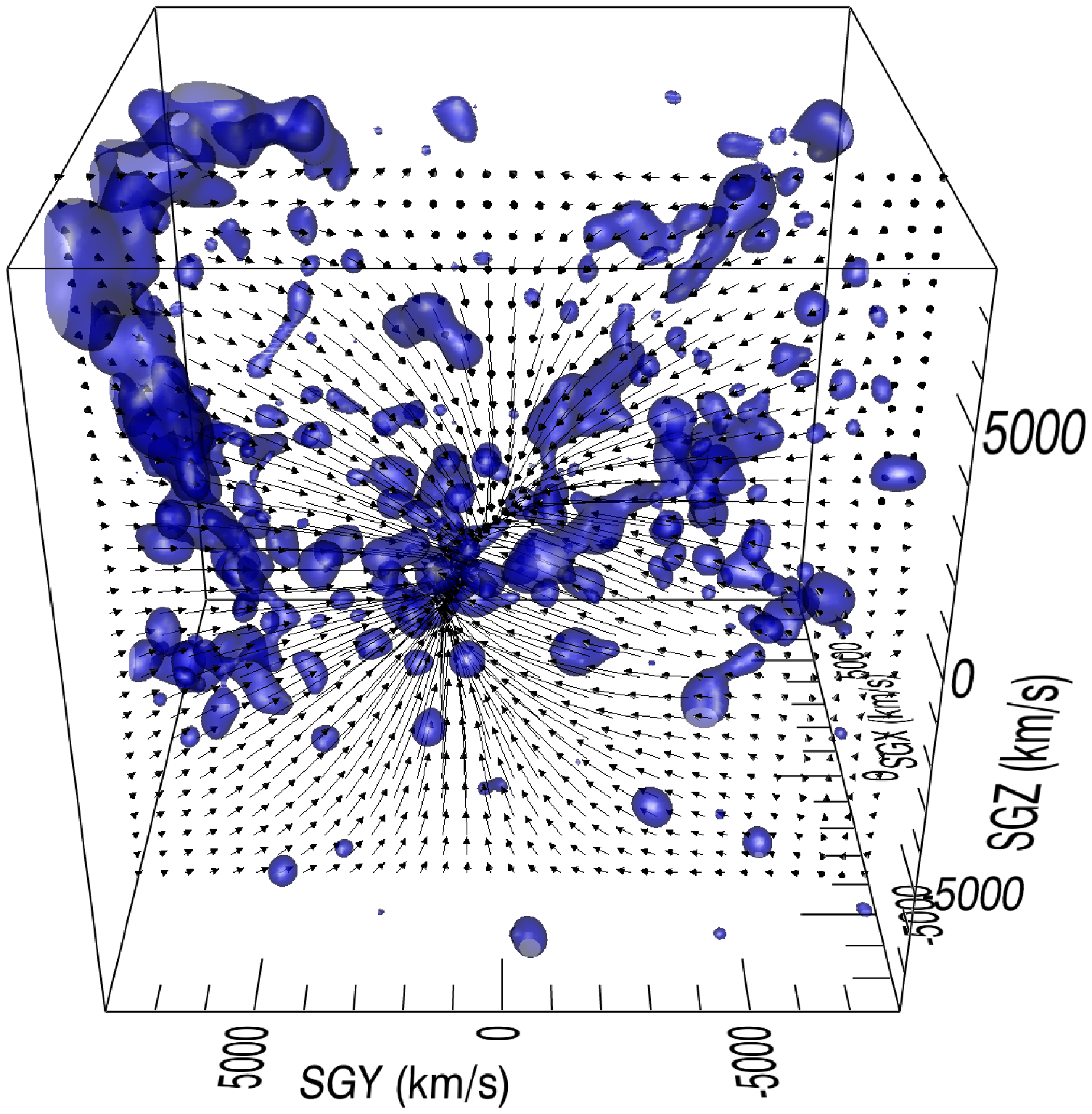} 
\vspace{-.2 cm}
 \caption{Visualization of the {\sl Cosmicflows-1} reconstructed velocity field on the
 Supergalactic Plane SGZ=0 (left) and on the SGY-SGZ plane going through the Centaurus Cluster (right).}
   \label{tomo}
\end{center}
\end{figure}

\begin{figure}[ht]
\begin{center}
 \includegraphics[width=0.905\textwidth]{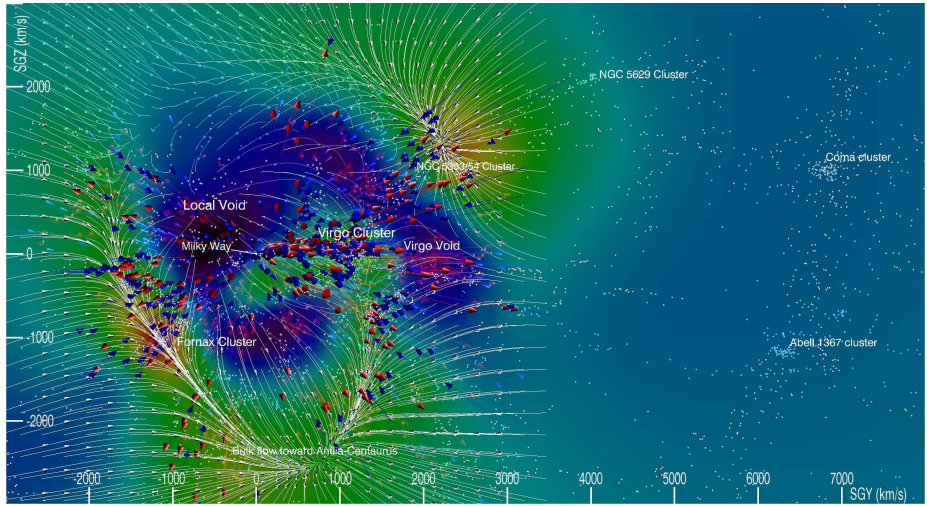} 
\vspace{-.15 cm}
 \caption{Visualization of structures and cosmic flows in the SGX=0 plane}
   \label{local2coma}
\end{center}
\end{figure}

One of the most striking feature in the structure of the Local Universe is the dense concentration of galaxies
identified by \cite[de Vaucouleurs (1956)]{supergalacticplane} that in large measure defines the equator of the
Supergalactic coordinate system.
Galaxies in this plane, at SGZ=0, are displayed in 
Fig.\,\ref{slices} (left). The Milky Way, located at the center,
belongs to the Local Supercluster whose main component is the Virgo Cluster. A dense filament connects Virgo to the region of
the Centaurus Cluster, as can be further appreciated in the other projection shown in Fig.\,\ref{slices} (right). At a distance similar
to Virgo Cluster but at negative SGY lies the Fornax Cluster. At 7000 km/s along the positive SGY axis lies the Great Wall
which includes the Coma Cluster. At intermediate distances on the scale considered here, we
find three major concentrations labelled Pavo-Indus, Perseus-Pisces, and the Southern Wall. In terms of cosmography, another striking feature
is the presence of voids with dimensions as large as 5000 km/s. These two maps are affected with the growing incompletness with distance. This is corrected in the three-dimensional map offered in Fig.\,\ref{widget}, which illustrate the
preponderance of the Great Wall as the most important structure of the Local Universe, and the relatively low importance of our
Local Supercluster.
Visualizations of the reconstructed velocity field in the nearby Universe are presented in Fig.\,\ref{tomo} together with a high-density
isosurface reconstructed from the {\sl V8k} catalog. From these maps, we can infer the presence
of a major attractor in the region of the Centaurus Cluster. 
Cosmic flows visualized as streamlines seeded in the SGX=0 plane are displayed in Fig.\,\ref{local2coma}.
This map also shows the underlying density field colored from black (underdense) to red (overdense),
the {\sl V8k} galaxies, and
the {\sl Cosmicflows-1} galaxies with radial peculiar velocities attached (blue and red arrows for inward and outward
moving galaxies, respectively).
This map reveals the presence of the very underdense ``Local Void" in the immediate vicinity of the Milky
Way. This void is seen in both the distribution of galaxies and in the reconstructed source field. The map illustrates the
clearing from the voids in the form of outbound flows, as already reported in
\cite[Tully \etal\ (2008)]{2008ApJ...676..184T}. The map also reveals that the Milky Way is caught in a flow toward the
Virgo Cluster, associated mostly with the Virgo gravitational infall. Other attractors and convergent flows seen
in this map are discussed in \cite[Courtois \etal\ (in prep)]{inprep}.

%\vspace{-.9 cm}
\section{Conclusions and perspectives}

The comparative visualization of three-dimensional structures and flows enrich the study of the cosmography of
the Local Universe. A more detailed discussion will be presented in \cite[Courtois \etal\ (in preparation)]{inprep}.
New observations being currently aggregated in the {\sl Cosmicflows-2} Catalog will provide distances up to 10,000 km/s. Reconstructions will extend our vision of cosmography to even larger scales.

\end{document}